# Tuning Optical Properties of Metamaterials by Mie Scattering for Efficient Sub-ambient Daytime Radiative Cooling

Shiwen Wu[1], Ruda Jian[1], Lyu Zhou[1], Siyu Tian[1], Tengfei Luo[2], Shuang Cui[1], Bo Zhao[3], Guoping Xiong[1, *]

[1]Department of Mechanical Engineering, The University of Texas at Dallas, Richardson, TX 75080, United States.

[2]Department of Aerospace and Mechanical Engineering, University of Notre Dame, Notre Dame, Indiana 46556, USA.

[3]Department of Mechanical Engineering, University of Houston, Houston, TX 77004, United States.

**\* Corresponding Authors**

E-mail: guoping.xiong@utdallas.edu (Guoping Xiong)

**Abstract:**

The management of the abundant eggshell biowaste produced worldwide has become a problematic issue due to the generated odor and microorganisms after directly disposing eggshell biowaste in landfills. Herein, we propose a novel method to convert the hazardous eggshell biowaste to valuable resources for energy management applications. Eggshell-based films are fabricated by embedding eggshell powders into polymer matrix to achieve highly efficient sub-ambient daytime radiative cooling. Benefiting from the Mie scattering of eggshell particles/air pores in the solar spectrum and strong emission of eggshell in the mid-infrared (mid-IR) range, the eggshell-based films present high reflection of 0.96 in the solar spectrum and high emission of 0.95 in the mid-IR range, resulting in significant average temperature drop of 5°C and 12°C below the ambient temperature during daytime and nighttime, respectively. Moreover, the eggshell-based films exhibit excellent flexibility and self-cleaning properties, which are beneficial for practical long-term outdoor applications. Our proposed design provides a novel means for an environmentally friendly and sustainable management of the eggshell biowaste.

# 1. Introduction

Eggs have become one of the most important food resources in world-wide feeding, because they provide essential nutrients for human body[1]. In 2020, the annual global egg production reached $8.7\times10^7$ tons and this value is still increasing rapidly[2,3]. However, enormous amount of eggshell is also generated from the daily consumption of eggs and is generally considered as a worthless, hazardous biowaste according to European Union regulations[6]. Most of the eggshell biowaste is commonly abandoned and directly disposed in landfills, resulting in gradual propagation of harmful odor and microorganisms[4,5]. Converting waste to valuable and useful resources has become a mainstreaming trend in 21st century according to the 3R principles (i.e., reduce, reuse and recycle) for promoting the Circular Economy[7]. Therefore, finding an effective route to manage the eggshell biowaste in an environmentally friendly and sustainable manner is highly desirable.

Recently, passive daytime radiative cooling has attracted tremendous attention due to its passive nature and potential to operate during the daytime[8–11]. The outer space with an ultralow temperature of approximately 3 K performs as a perfect heat sink for heat transfer processes. The coincident overlapping between the atmospheric transparency window on the earth (i.e., 8-13 μm) and the spectral peak of blackbody radiation at 300 K provides objects on the earth an efficient way to eliminate heat and thus passively cool down their temperature through thermal radiative emissions into the outer space[12,13]. Given that light absorption within the solar spectrum is minimized simultaneously, passive cooling of surfaces below ambient temperature can be achieved even during the daytime without external energy input[14]. The main component of eggshell is calcium carbonate which possesses abundant absorption peaks and thus high emission in the mid-infrared (mid-IR) range[15,16]. Therefore, eggshell biowaste is highly possible to act as an effective component in radiative cooling materials.

Herein, we report the design of eggshell biowaste-derived films with excellent flexibility and self-cleaning functions to achieve sub-ambient daytime radiative cooling. Through optimizing experimental parameters, the sizes of eggshell particles and air pores in the polymer matrix are adjusted to a sub-micron range to efficiently scatter sunlight by Mie scattering. Meanwhile, strong absorption of eggshell powders originated from the stretching and bending of C-O and Ca-O bonds provides high emission in the mid-IR range. Consequently, high reflection of 0.96 in the solar spectrum and high emission of 0.95 in the mid-IR range are achieved by the eggshell-based films, which are comparable to or even higher than those of the state-of-the-art radiative cooling materials. Outdoor field tests show that average temperature drop of 5°C and 12°C below the ambient temperature during daytime and nighttime can be achieved, respectively. Moreover, the outstanding flexibility and water-repellent properties enable the eggshell-based films to meet complex outdoor conditions without deteriorating the radiative cooling performance. Our design provides an effective way to convert the eggshell biowaste to environmentally friendly functional materials.

## 2. Experimental

*2.1. Preparation of eggshell powders*

Eggshells were collected from consumed eggs purchased from local supermarkets. Firstly, 10 g eggshell GO was added into 100 ml 6% $H_2O_2$ solution and heated at 90°C for 1 h to remove the impurities. After being rinsed by water for several times and dried at 80°C, the eggshells were broken down to eggshell powders by a vertical planetary ball mill machine (Biobase Bkbm-V0.4).

Zirconium dioxide balls and containers were used in the ball-milling process, and the milling speed was fixed at 700 rpm for 24 h.

*2.2. Preparation of eggshell-based polymethylpentene (E-TPX) films*

5.4 g ball-milled eggshell powders and 1.2 ml KH550 (Sigma Aldrich) were added into 40 mL cyclohexane (Sigma Aldrich) to prepare eggshell/cyclohexane suspensions. Then, 2.23 g polymethylpentene (TPX) (Sigma Aldrich) were added and dissolved in the cyclohexane solvent at 65°C. Subsequently, the prepared eggshell/TPX solutions were poured into a tap-casting mold and frozen at -60°C. E-TPX films were obtained after a following freeze-drying process for 48 h.

*2.3. Characterization*

Reflectivity spectra in the solar spectrum and emissivity spectra in the mid-IR range were measured by a UV–vis NIR spectrometer (Agilent Cary 5000) and Bruker VERTTEX 70 FTIR equipped with integrating spheres, respectively. The FTIR spectra of the eggshell powders were recorded using a FTIR spectrometer (Agilent 660-2). The morphology and microstructures of the samples were characterized using a field-emission scanning electron microscope (SEM, Supra 40, Zeiss).

## 3. Results and discussion

Figure 1 schematically shows the fabrication process of the E-TPX films. Here TPX is employed as the polymer matrix because of its excellent transmittance in both solar spectrum and infrared range[17,18]. Briefly, grinded eggshell powders were initially dispersed in cyclohexane, and then TPX was added and dissolved into the cyclohexane solvents to obtain eggshell/TPX solutions. The amount of eggshell powders and TPX was optimized to achieve efficient passive daytime radiative cooling (Details shown in Supplementary Section S1). Subsequently, the eggshell/TPX solution was tape-casted with a fixed thickness of 600 μm and frozen at -60°C. After the sublimation of frozen cyclohexane solvent during the freeze-drying process, E-TPX films containing a large amount of embedded air pores and eggshell particles were obtained.

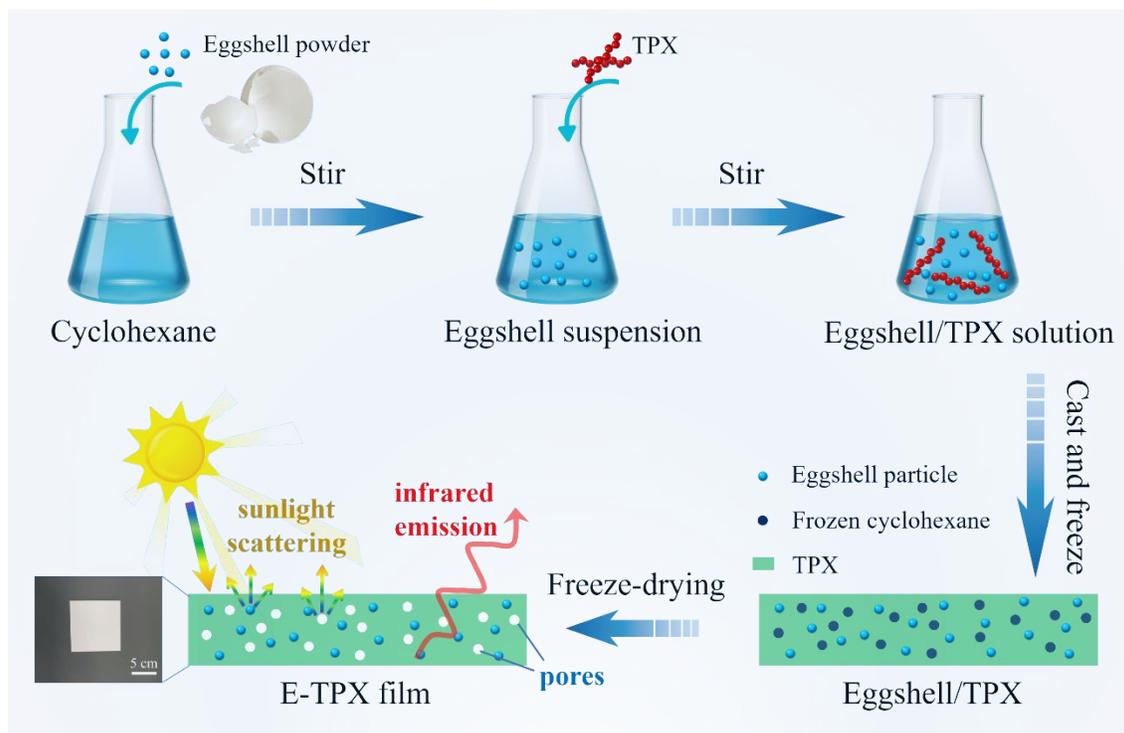

Figure 1. Schematic showing the fabrication process of E-TPX films.

The E-TPX film exhibits outstanding mechanical performance. As shown in Figure 2a, an E-TPX film with a width of 1 cm can hold up to 500-g weight. Moreover, the free-standing E-TPX film can be easily recovered to the original shape after being twisted, bended or curved to a spiral shape (Figure 2b), demonstrating its excellent flexibility which is highly desirable to cool surfaces with complex structures for wider applications. Figure 2c presents the cross-sectional SEM images of the E-TPX films at different magnifications. Eggshell particles with diameters around several hundreds of nanometers are uniformly distributed in the TPX matrix. Abundant air pores can also be observed in the TPX networks, with the size optimized to a sub-micron range by adjusting the ratio of TPX to cyclohexane in the fabrication process (Supplementary Figure S4). The eggshell particles and air pores with sub-micron sizes concurrently increase the scattering cross sections in the visible and near-IR ranges due to Mie scattering[19], leading to high reflection in the solar spectrum (Supplementary Figure S5). Meanwhile, the chemical structures of eggshell powders are characterized by FTIR and shown in Figure 2d. The FTIR spectrum of eggshell powders exhibits strong absorption peaks in the mid-IR range. The peaks at 1398 and 873 cm$^{-1}$ correspond to the C-O stretching and bending of the main component, $CaCO_3$, in eggshell powders, while the sharp peak at 710 cm$^{-1}$ represents the Ca-O bond[16]. These absorption peaks of eggshell powders enable strong emission of the E-TPX films in the mid-IR range. Consequently, the E-TPX films exhibit exceptional optical properties with weighted average reflection of 96% in the solar spectrum and emission of 95% in the mid-IR range, which are comparable to or even higher than those of the state-of-the-art passive daytime radiative cooling materials[10,11,20–23].

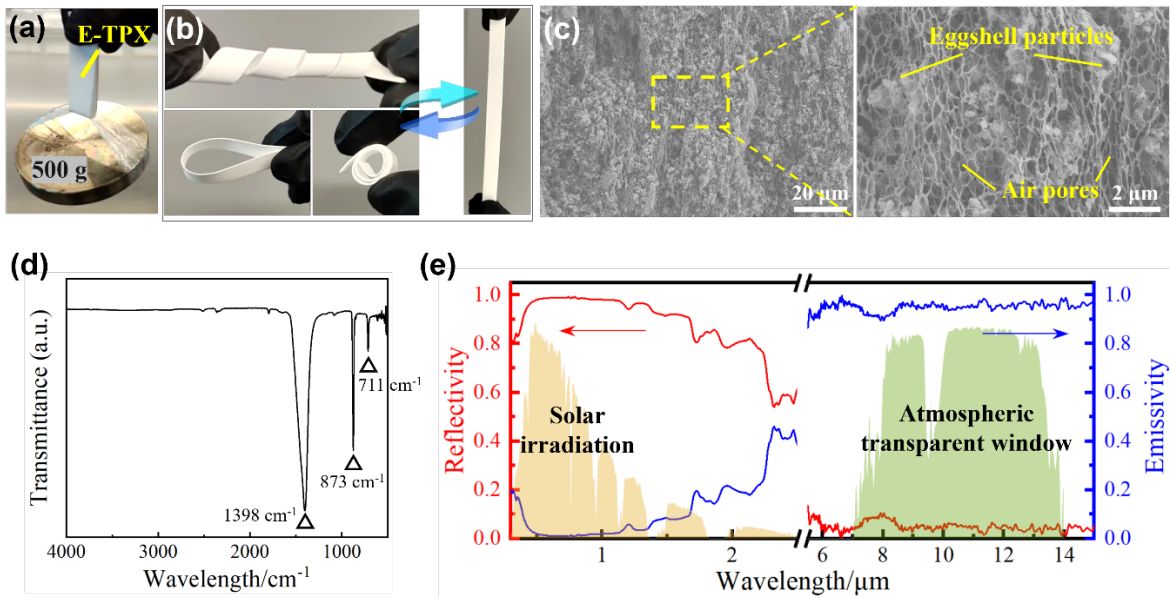

Figure 2. (a) Photograph showing that an E-TPX film with a 1-cm width can hold 500-g weight. (b) Photographs showing that the E-TPX film can be recovered to the original shape after being twisted, bended or curved to a spiral shape. (c) The cross-sectional SEM images of the E-TPX films. (d) FTIR spectrum of the eggshell powders. (e) Reflectivity and emissivity spectra of the E-TPX films in the solar spectrum and mid-infrared regime.

The radiative cooling performance of the E-TPX films was evaluated on sunny days in Richardson (96.73°W, 32.95°N, Texas, U. S.). Figures 3a and 3b present our customized experimental setup for radiative cooling field tests. The radiative cooling samples were placed on the top of a highly insulating vacuum insulation panel (VIP, with a thermal conductivity of ~ 2 mW m$^{-1}$ K$^{-1}$), sitting inside an expanded polystyrene foam (EPS, with a thermal conductivity of ~ 30-40 mW m$^{-1}$ K$^{-1}$) box to minimize the parasitic heat gains to the back side of the samples. The EPS box was further covered by a 10-μm-thick polyethylene film to reduce convective heat loss and improve thermal isolation. Two k-type thermocouples, one placed beneath the testing films and the other one placed in an open aluminum foil pocket, were used to measure the achieved sub-ambient cooling temperatures by the testing samples and ambient temperatures, respectively. Benefiting from the high solar reflection and high infrared emission, significant sub-ambient cooling effect was achieved by the E-TPX films during a 48-h continuous test (Figure 3c). To further reveal the sub-ambient cooling performance of the E-TPX films, temperature differences ($\Delta T$) between the ambient and the E-TPX films during the daytime and nighttime are calculated and plotted in Figure 3d. The E-TPX films maintained an average temperature drop of 5°C and 12°C below the ambient temperature around noon and during nighttime, respectively. Another outdoor field test was performed in Reno (119.81°W, 39.53°N, Nevada, U. S.) with a completely different weather condition, where an average temperature drop of 6°C and 13°C below the ambient temperature around noon and during nighttime were observed (Supplementary Figure S6), respectively, indicating the outstanding sub-ambient radiative cooling capability of the E-TPX films.

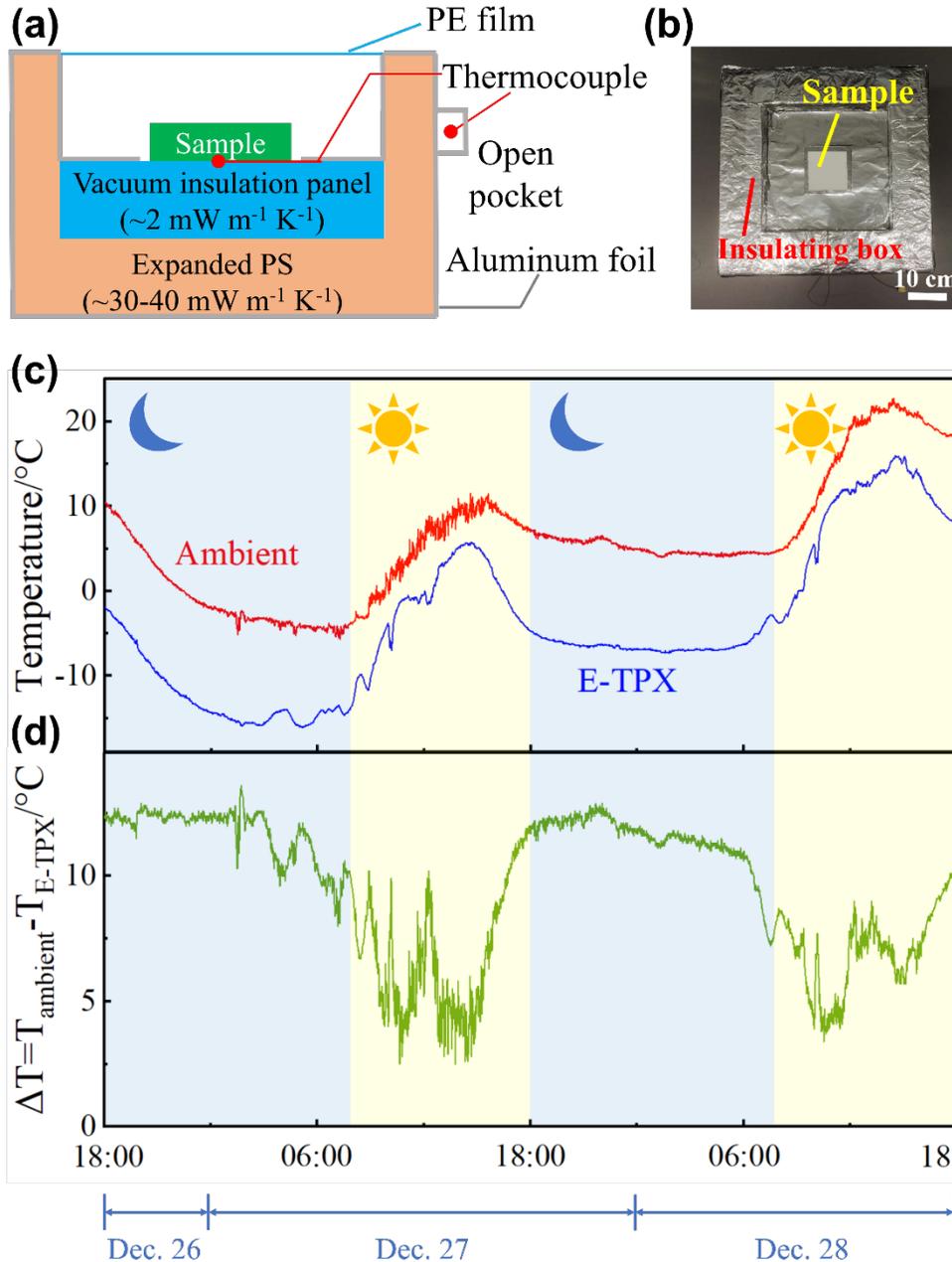

Figure 3. (a) Schematic and (b) photograph of our experimental setup for radiative cooling field tests. (c) Measured temperatures of the ambient and the E-TPX film as a function of time when exposed on sunny days in Richardson, Texas, U.S. (18:00, December 26, 2022 to 18:00, December 28, 2022). (d) Temperature differences (ΔT) between the ambient and the E-TPX films.

Surfaces of radiative cooling materials inevitably suffer from rain flushing and dust contamination when exposed to outdoor environments, which may result in severe deterioration of cooling performance. Our designed E-TPX films exhibit superhydrophobic wettability with a water contact angle of 150.9° (Figure 4a) and can repel a wide range of possible liquid contaminants (e.g., coke, milk, coffee, soy source, and juice) in daily life (Figure 4b). The superhydrophobicity and liquid-repellent properties protect E-TPX films from being damaged by rains and liquid contaminants. Moreover, self-cleaning performance of the E-TPX films is also demonstrated. Ink droplet and clay were placed on the surface of the E-TPX film as representatives of liquid and solid contaminants, respectively. These contaminants can be easily removed from the E-TPX film by rinsing with water droplets. In this manner, the E-TPX films can repel water and clean the dirt contaminations in rainy days, indicating the outstanding durability for outdoor applications.

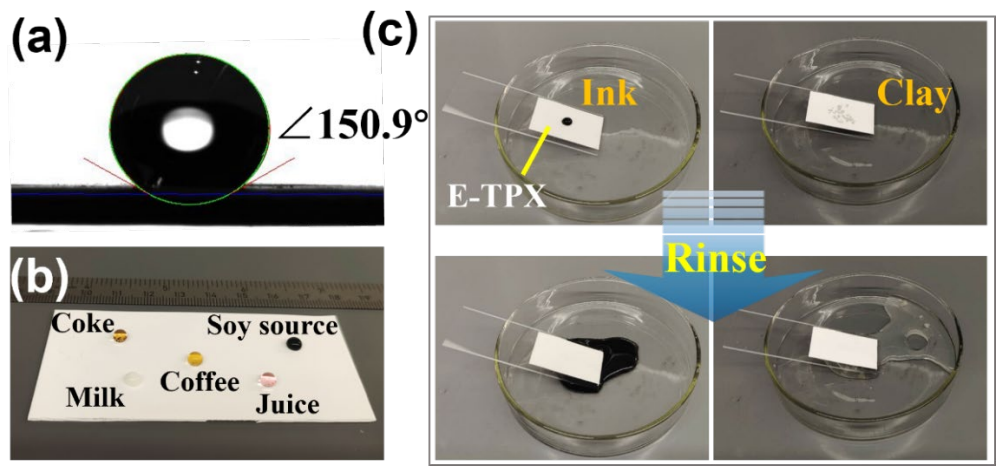

Figure 4. (a) Water contact angle on the E-TPX film. (b) Photograph showing the droplets of coke, milk, coffee, soy source, and juice placed on the E-TPX film. (c) Photographs showing the self-cleaning capability of the E-TPX films. Ink and clay can be easily removed from the E-TPX film by water rinsing.

## 4. Conclusions

In summary, we report eggshell biowaste-derived films for efficient sub-ambient daytime radiative cooling. The as-prepared E-TPX films exhibit high reflection of 0.96 in the solar spectrum due to the Mie scattering of embedded eggshell particles and abundant air pores in the TPX matrix, and high emission of 0.95 in the mid-IR range due to the strong infrared absorption of eggshell. Consequently, the radiative cooling material achieves sub-ambient cooling effect with an average temperature drop of 5°C and 12°C below the ambient temperature during daytime and nighttime, respectively. Moreover, the E-TPX films show excellent flexibility and self-cleaning properties, which is crucial for long-term outdoor applications. The design of E-TPX films provides a novel means for the sustainable and environmentally friendly management of eggshell biowastes.

## Acknowledgements

G.X. thanks the University of Texas at Dallas startup fund and the support from the NSF (Grant No. CBET-1937949). T.L. thanks the support from the NSF (Grant No. CBET-1937923). The authors thank Bobby Lee Collins for the assistance involved in preparing the outdoor field test setup.

# Supplementary Materials for

# Tuning Optical Properties of Metamaterials by Mie Scattering for Efficient Sub-ambient Daytime Radiative Cooling

Shiwen Wu[1], Ruda Jian[1], Lyu Zhou[1], Siyu Tian[1], Tengfei Luo[2], Shuang Cui[1], Bo Zhao[3], Guoping Xiong[1, *]

[1]Department of Mechanical Engineering, The University of Texas at Dallas, Richardson, TX 75080, United States.

[2]Department of Aerospace and Mechanical Engineering, University of Notre Dame, Notre Dame, Indiana 46556, USA.

[3]Department of Mechanical Engineering, University of Houston, Houston, TX 77004, United States.

* Corresponding authors: Guoping Xiong: guoping.xiong@utdallas.edu

# Section S1 Optimization on experimental parameters

Minimizing absorption in the solar spectrum is of great importance to achieving high-performance daytime radiative cooling. The E-TPX films rely on Mie scattering of the embedded eggshell particles and air pores which is sensitive to the particle/pore size[19]. Here we adjust the geometric sizes of the eggshell particles and air pores through changing the amount of added eggshell powders and TPX.

Firstly, the size of air pores is optimized by changing the ratio of TPX/cyclohexane. Porous TPX films are fabricated by the same tape-casting and freeze-drying processes. We denote the TPX films prepared by dissolving 0.45 g, 1.12 g, 2.23 g, and 3.35 g TPX into 40 mL cyclohexane as T-1, T-2, T-3, and T-4 samples, respectively. The reflectivity of porous TPX films in the solar spectrum increases with the amount of added TPX (Figure S1). However, When the concentration of TPX is too high, the eggshell/TPX solution becomes too viscose to disperse eggshell particles uniformly. Therefore, the amount of added TPX is fixed as 2.23 g/40 mL cyclohexane.

Then, different E-TPX films are prepared by adding 1.8 g, 3.6 g, 5.4 g, and 7.2 g eggshell powders into 40 mL cyclohexane, which are denoted as E-TPX-1, E-TPX-2, E-TPX-3, and E-TPX-4, respectively. The amount of added eggshell powders is optimized to be 5.4 g (i.e., E-TPX-3) according to the measured optical properties (Figure S2), because the emissivity of E-TPX films decreases significantly with a lower content of eggshell, while higher content of eggshell leads to lower solar reflectivity due to the absorption in UV regime and aggregation of eggshell particles in TPX matrix (Figure S3).

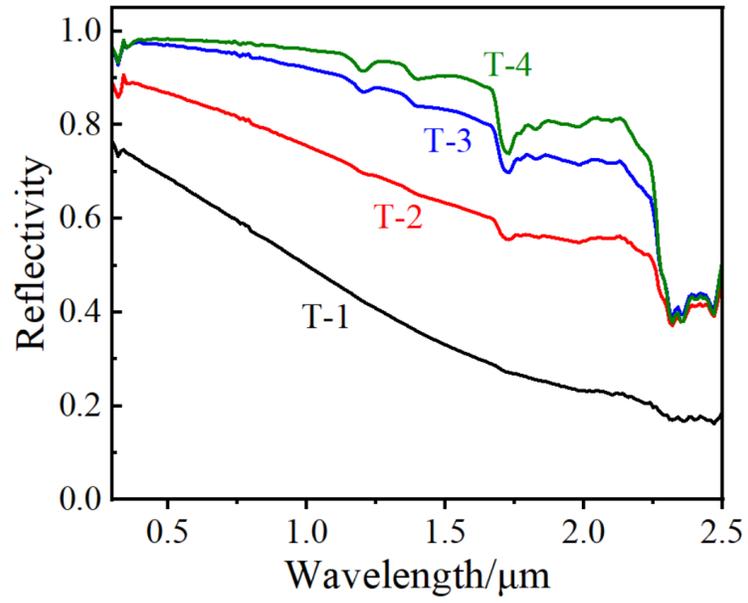

Figure S1. Reflectivity spectra of the porous TPX films in the solar spectrum.

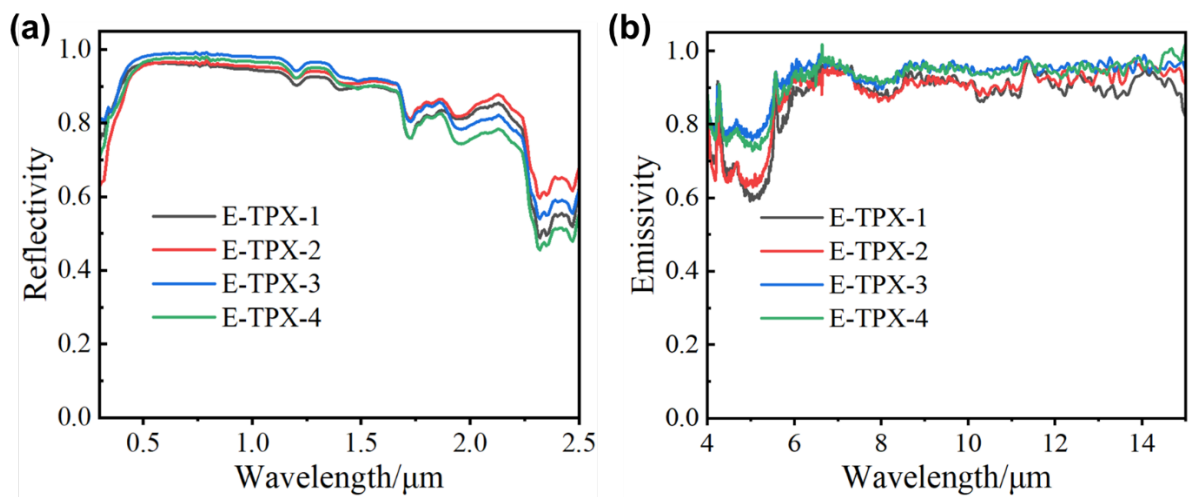

Figure S2. Reflectivity and emissivity spectra of the E-TPX films in the solar spectrum and mid-IR regime.

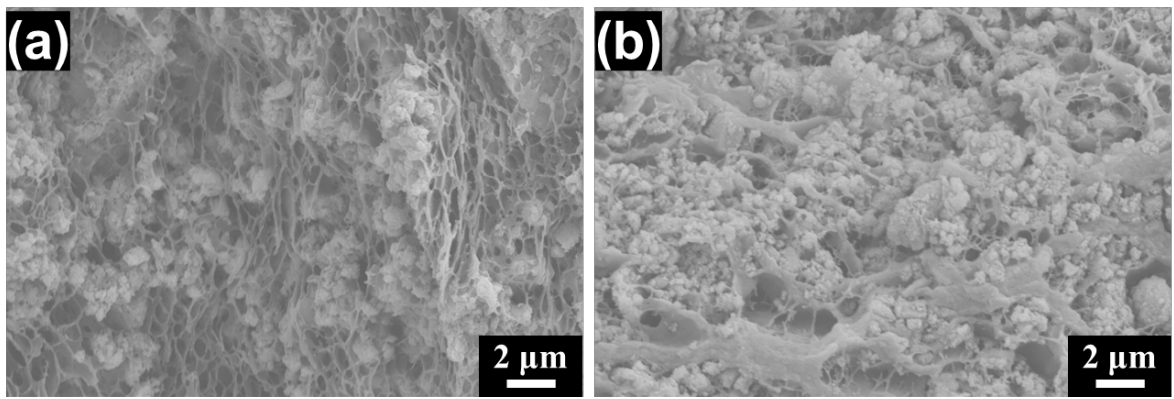

Figure S3. The cross-sectional SEM images of the E-TPX films: (a) E-TPX-3 and (b) E-TPX-4.

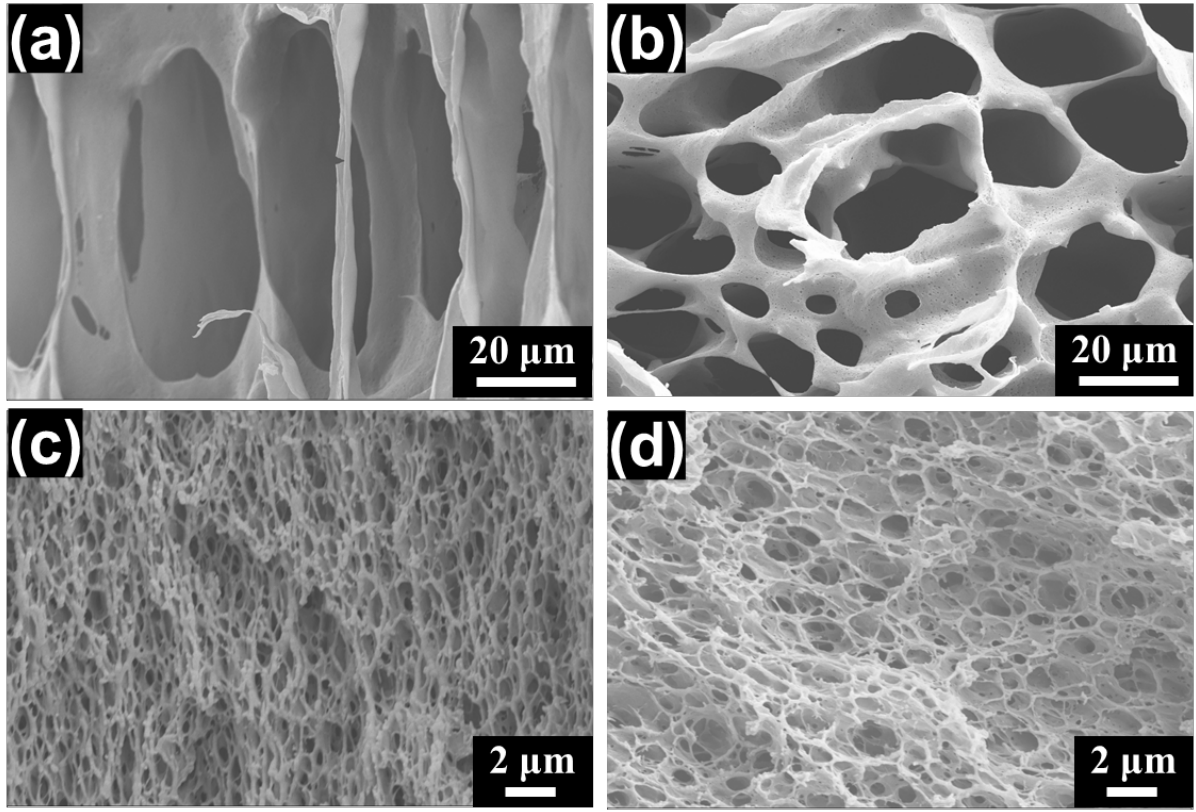

Figure S4. The cross-sectional SEM images of the porous TPX films: (a) T-1, (b) T-2, (c) T-3, and (d) T-4.

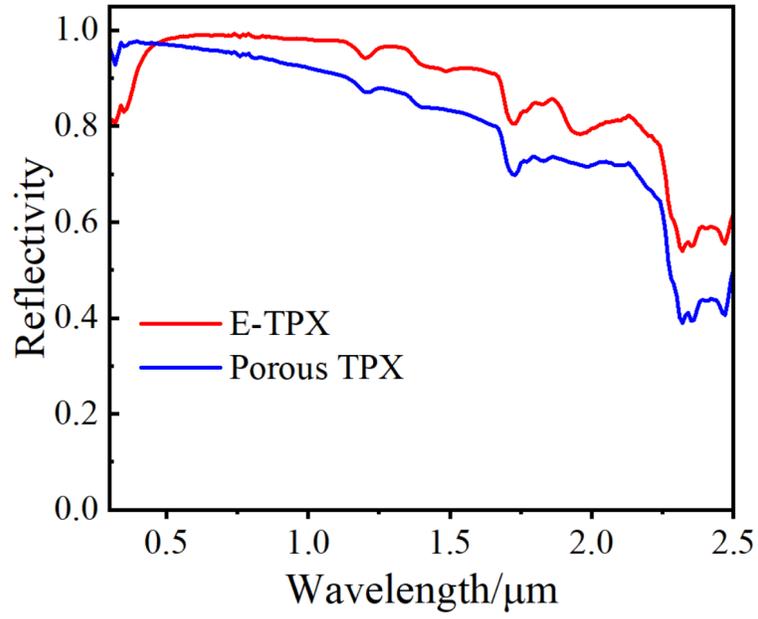

Figure S5. Reflectivity spectra of the porous TPX and E-TPX films in the solar spectrum. The existence of eggshell particles further increases the solar reflection.

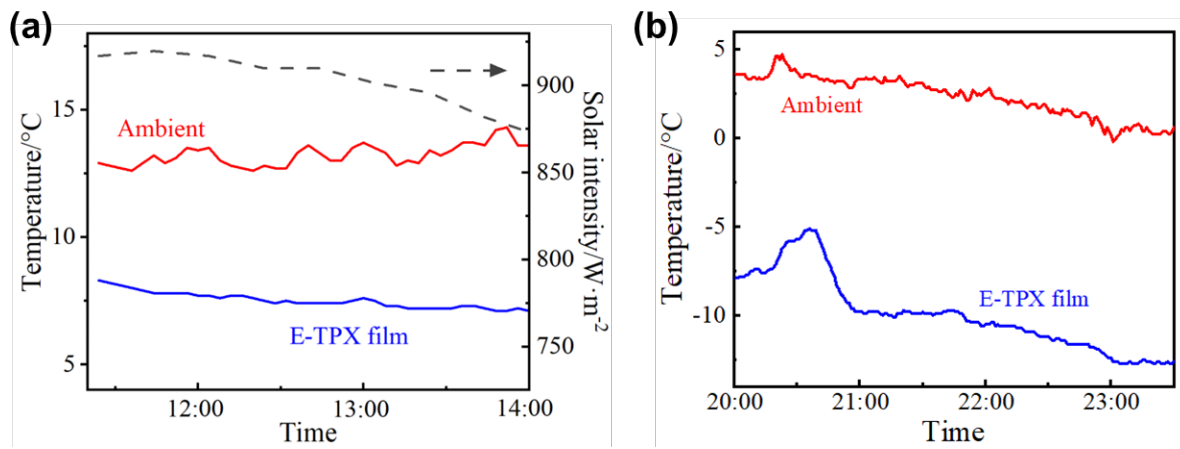

Figure S6. Measured temperatures of the ambient and the E-TPX film as functions of time when exposed on sunny days in Reno, Nevada, U.S.